\documentclass[lettersize,journal]{IEEEtran}

\usepackage{amsmath,amsfonts}
\usepackage{algorithmic}
\usepackage{algorithm}
\usepackage{array}
\usepackage[caption=false,font=normalsize,labelfont=sf,textfont=sf]{subfig}
\usepackage{textcomp}
\usepackage{stfloats}
\usepackage{url}
\usepackage{verbatim}
\usepackage{graphicx}
\usepackage{cite}
\usepackage{multirow}

\begin{document}

\title{LEMDA: A Novel Feature Engineering Method \\for Intrusion Detection in IoT Systems}

\author{Ali~Ghubaish,~\IEEEmembership{Graduate~Student~Member,~IEEE,}								    Zebo~Yang,~\IEEEmembership{Graduate~Student~Member,~IEEE,}
Aiman~Erbad,~\IEEEmembership{Senior~Member,~IEEE,}
            and~Raj~Jain,~\IEEEmembership{Life~Fellow,~IEEE}
        % <-this % stops a space
\thanks{Ali Ghubaish is with Washington University in Saint Louis, St. Louis, MO 63130 USA (email: aghubaish@wustl.edu).}
\thanks{Zebo Yang is with Washington University in Saint Louis, St. Louis, MO 63130 USA (email: zebo@wustl.edu).}
\thanks{Aiman Erbad is with College of Science and Engineering, Hamad Bin Khalifa University, Qatar (email: aerbad@hbku.edu.qa).}
\thanks{Raj Jain is with Washington University in Saint Louis, St. Louis, MO 63130 USA (email: jain@wustl.edu).}\\
\thanks{Manuscript received October 10, 2022.}}

\markboth{Journal of \LaTeX\ Class Files,~Vol.~14, No.~8, August~2021}%
{Shell \MakeLowercase{\textit{et al.}}: A Sample Article Using IEEEtran.cls for IEEE Journals}


\maketitle
%-------------------------------------------------------------------------------
\begin{abstract}
%-------------------------------------------------------------------------------

Intrusion detection systems (IDS) for the Internet of Things (IoT) systems can use AI-based models to ensure secure communications. IoT systems tend to have many connected devices producing massive amounts of data with high dimensionality, which requires complex models. Complex models have notorious problems such as overfitting, low interpretability, and high computational complexity. Adding model complexity penalty (i.e., regularization) can ease overfitting, but it barely helps interpretability and computational efficiency. Feature engineering can solve these issues; hence, it has become critical for IDS in large-scale IoT systems to reduce the size and dimensionality of data, resulting in less complex models with excellent performance, smaller data storage, and fast detection. This paper proposes a new feature engineering method called \textbf{LEMDA} (\textbf{L}ight feature \textbf{E}ngineering based on the \textbf{M}ean \textbf{D}ecrease in \textbf{A}ccuracy). LEMDA applies exponential decay and an optional sensitivity factor to select and create the most informative features. The proposed method has been evaluated and compared to other feature engineering methods using three IoT datasets and four AI/ML models. The results show that LEMDA improves the $F_1$ score performance of all the IDS models by an average of 34\% and reduces the average training and detection times in most cases.  
\end{abstract}


%-------------------------------------------------------------------------------
\section{Introduction}
%-------------------------------------------------------------------------------

\IEEEPARstart{I}{nternet} of Things (IoT) plays a signiﬁcant role in 21st-century technologies. According to  Schiller et al.~\cite{IoTAnalytics2022}, the number of IoT devices is expected to jump from 31 billion in 2022 to 75 billion in 2025. Even with COVID-19 affecting supply and demand, resulting in a global chip shortage, these numbers were only slightly less, from 11.7 to 11.3, than forecast in 2020. This is because it integrates into different application domains of our daily lives, from Industrial IoT (IIoT) to critical Internet of Medical Things (IoMT). This increase in IoT systems will be accelerated in the 5G era as the network infrastructure enables large-scale cellular IoT (e.g., massive IoT). Due to the vast number of IoT devices in most IoT networks, these networks generate massive high-dimensional datasets, causing feature explosion~\cite{Lincbi}.

Securing IoT systems is crucial due to the sensitive nature of the data produced by these sensors and devices (e.g., the medical data generated by IoMT systems). One of the solutions to secure these systems is by using large-scale intrusion detection systems (IDS), which can handle the massive and high-dimensional data generated by these systems \cite{Thakkar,Vijayanand}. However, fitting these colossal datasets requires complex machine learning (ML) models, which brings challenges of overfitting, low interpretability, and high computational complexity. Besides, with high dimensional data, ML models for IDS require powerful devices and significant training time to detect attacks accurately. Thus, it is non-trivial to avoid the high complexity of data. Reducing complexity in data before modeling has been a hot topic in recent years \cite{CZ23}. Feature engineering is one popular method of such techniques. Feature engineering methods help select the best features for these models, expediting the processes of finding the optimal hyperparameters for the IDS models. We use “feature engineering” and “feature reduction” interchangeably in the rest of this paper to describe the methods that reduce the datasets’ dimensionality.

Most informative features can be selected using feature engineering or dimension reduction techniques, such as feature selection and feature extraction. Feature selection techniques help simplify complex models by reducing the dimensionality of the dataset (number of features), which avoids over-fitting and results in less training time and storage space. Only the most important features are retained after feature selection. On the other hand, feature extraction techniques, such as principal component analysis (PCA), create new features that preserve the data’s variance based on existing features.

Feature selection techniques are divided into four categories: ﬁlter, wrapper, embedded, and hybrid \cite{Yang21}. Filter methods are fast, but they may fail to select the most informative features and thus lead to low accuracy in ML models. Wrapper methods, like recursive feature elimination (RFE) and forward feature selection (FFS), are effective in selecting informative features but are slow and susceptible to overfitting. Embedded methods, such as the mean decrease in impurity (MDI) and the mean decrease in accuracy (MDA), provide a tradeoff between accuracy and speed, thus providing balanced results between filter and wrapper methods. Finally, hybrid methods are a mix of two or more of these methods (e.g., \cite{Shaban, Kamarudin}) and feature extraction such as PCA (e.g., \cite{Li20}). However, these methods are usually designed for specific datasets or models. More about hybrid methods can be found in \cite{Shekhawat21}.

This paper proposes a new feature engineering method for supervised ML-based IDS in IoT systems called Light feature Engineering based on the Mean Decrease in Accuracy (LEMDA). It takes the benefits of both feature selection and extraction methods and reduces their drawbacks. LEMDA is based on embedded methods and achieves a better tradeoff between performance and speed.

Our method consists of two parts: 1) creating a list of the most informative features using the MDA method 2) creating a new feature from the first feature (the most informative one) in that list. The new feature is created using the weighted exponential decay formula (WEDF) technique. In addition, in cases where the most informative feature is categorical, we utilize the Sensitivity Factor (SF) to complement the WEDF method for creating a new feature. This happens, for example, when most attacks are passive, e.g., sniffing. WEDF and SF optimize the relationship between the values in the most informative feature and the samples’ classes, as shown in the Evaluation section of this paper.

LEMDA is a general feature engineering method using AI-based models for the supervised ML-based IDS in IoT systems. We demonstrate the effectiveness of our method by using three different datasets and comparing three different ML models using three different metrics. The evaluation results show the outstanding performance of LEMDA in IDS, with high accuracy and low detection time. The main contributions of this paper can be summarized as follows:
\begin{itemize}
\item We present LEMDA, a novel feature engineering method designed for supervised ML-based IDS in IoT.
\item We show that the WEDF, when added to MDA, significantly improves the performance.
\item We develop an add-on technique, SF, to enhance performance in cases where the most informative feature is categorical, which happens when most attacks are passive, e.g., sniffing.
\item We have designed LEMDA to be compatible with a wide range of supervised ML models, such as random forest (RF), allowing straightforward implementation without hyperparameter tuning.
\item We evaluate the performance of our method using three different datasets of different sizes, four different AI/ML models, and three different metrics – a total of 36 combinations.
\item We empirically show that LEMDA improves the IDS performance by 34\% in all cases and significantly reduces training and detection times in most cases.
\end{itemize}
The remainder of this article is organized as follows. A brief background of the commonly used feature engineering methods and the related work is provided in Sections II and III, respectively. In Section IV, we present our proposed method. The experimental methodology and results are shown in Sections V and VI. Finally, we conclude this paper in Section VII.

%-------------------------------------------------------------------------------
\section{Background}
\label{section:bg}
%-------------------------------------------------------------------------------
This section presents a background of the most commonly used feature engineering methods.

\subsection{Categories of Feature Selection Techniques}
%-----------------------------------
The difference between the four common feature selection techniques – Filter, Wrapper, Embedded, and Hybrid – is briefly explained in this subsection. More detailed information can be found in \cite{Yang21} and \cite{Shekhawat21}.
\\
\textbf{1. Filter:}
In this technique, features are sorted based on their relevance. Then, a threshold is applied to select the features that have strong relevance. This results in a fast selection but may lead to low accuracy if the dataset distribution is not uniform and the features are highly correlated. The correlation coefficient method is an example of this technique. It measures the linear relationship between the features and selects the features with a correlation below a specific threshold.
\\
\textbf{2. Wrapper:}
 In this method, the features are selected by measuring the performance improvement for an ML model using a subset of the features. The subset with the highest improvement in the ML model is selected. This technique effectively selects informative features but is very slow since it is computationally expensive, and the complexity increases as the number of features increases. For example, the RFE method uses all the features at the beginning, recursively removes them, and then sorts them by their incremental improvement.
\\
\textbf{3. Embedded:}
Embedded techniques combine the advantages of filter and wrapper techniques by embedding feature selection within the ML model itself. However, this makes it less generic than filter and wrapper techniques. MDI and MDA methods are examples of this technique, which will be explained in detail in the next subsection.
\\
\textbf{4. Hybrid:}
In this technique, two or more filter and wrapper methods (e.g., \cite{Shaban}) are combined to select a subset of the features to take advantage of each method and avoid their disadvantages. It is similar to the embedded technique but is more generic.

\subsection{Existing Feature Engineering Methods}
%----------------------------------
We chose the embedded technique for comparison with our method among the four feature selection techniques, considering their balance between accuracy and selection time. Specifically, we delve into two embedded feature selection methods, MDI and MDA. Additionally, we introduce the PCA feature extraction technique as part of our comparison, as it is commonly used in similar studies.

By introducing the three techniques, this subsection aims to clarify the differences between our method and existing methods, which will be highlighted and compared in the results section.
\\
\textbf{1. MDI method:}

MDI, also called Gini importance, is based on RF and is used to calculate the importance of each feature based on the weighted sum of the actual decrease in impurity for each feature across all trees \cite{Farrukh}. The larger the MDI score, the more important the feature. Since IDS uses a binary classification model, labeling with normal and attack, the decrease in impurity ($I$) can be calculated using Eq. \ref{eqn:1}:
\begin{equation} 
\label{eqn:1}
I = G_{PE} - P_{LS}G_{LS} - P_{RS}G_{RS}
\end{equation}

Here, $G_{PE}$ is the parent Gini (G) impurity index, as shown in Eq. \ref{eqn:2}. $G_{LS}$ and $G_{RS}$ are G indices for the left and right splits from the parent node in the tree, and $P_{LS}$ and $P_{RS}$ are the proportions for each split from their parent node (i.e., $P_{LS}+P_{RS}=1$).
\begin{equation} 
\label{eqn:2}
G = \sum_{i=1}^{n_c} p_i(1-p_i)
\end{equation}

Here, $n_c$ is the number of classes, which in our case is 2, and $p_i$ is the ratio for the $i^{th}$ class. G equals 0.5 if the number of samples for each class is the same and 0 if only one class is found in the dataset. However, this method is known to be biased toward high cardinality features \cite{Ward21}.
\\
\textbf{2. MDA method:}

MDA is also called permutation importance (Perm) and is similar to MDI as both are based on RF \cite{Ward21}. This method requires a validation set to calculate the importance score for each feature ($f$). This score is the weighted difference between the model's prediction error rate for the validation set before and after the permutation of each feature $f$ across all the trees, as shown in Eq. \ref{eqn:3}:

\begin{equation}
\label{eqn:3}
 MDA \;\;\textnormal{score}_{f}=\frac{1}{n_t}\sum_{j=1}^{n_t}(sa_{jf}-sb_{jf})
 \end{equation}
 
Here $n_t$ is the number of trees, $sa_{jf}$ and $sb_{jf}$ are the scores after and before permutating feature $f$ in the $j^{th}$ tree, respectively. Similar to MDI, the larger the score, the more important the feature. In general, MDA can result in ignoring more irrelevant features than the MDI method. 
\\
\textbf{3. PCA method:}

PCA differs from the previous two methods since it creates new features from the original ones.  These new features are called principal components (PCs) that are uncorrelated and represented by a set of eigenvectors \cite{Shekhawat21}. These eigenvectors and their corresponding eigenvalues are calculated using a covariance matrix. The PCs are sorted in descending order based on their explained variance, where the first PC has the highest explained variance among all features. The explained variance for each PC ($var\_PC_{i}$) is the ratio of that PC's eigenvalue ($\lambda_i$) to the sum of all eigenvalues, as shown in Eq \ref{eqn:4}.

\begin{equation}
\label{eqn:4}
 var\_PC_{i}=\frac{\lambda_i}{\sum_{i=1}^{n_{pc}} \lambda_i}
 \end{equation}

The easiest and most effective way to set the required number of PCs ($n_{pc}$) with good performance is by setting a threshold to calculate the necessary number of PCs to get 95\%-99\% explained variance \cite{Vilsen21}. PCA improves model performance and is versatile to most ML models, but it is laborious to tune the threshold.

%-------------------------------------------------------------------------------
\section{Related Work}
\label{section:related_work}
%-------------------------------------------------------------------------------
The classification of IDS can be divided into five categories: network IDS (NIDS), host IDS (HIDS), protocol-based IDS (PIDS), application protocol-based IDS (APIDS), and hybrid IDS \cite{Swain21}. NIDS \cite{Nguyen22, Yu22} are designed to monitor the network traffic of all network communications and are usually centralized in one point of the system, such as the cloud. On the other hand, HIDS \cite{Lightbody22} only monitors the traffic of only one device. PIDS \cite{Mandal21, Zeeshan22} and APIDS are set up to monitor specific protocol connections, e.g., hypertext transfer protocol secure (HTTPS), and application-specific protocols, e.g., structured query language (SQL), respectively. Hybrid IDS integrates multiple IDS mentioned above to leverage each IDS type's strengths. 

Most of the IDS in the IoMT systems are NIDS since the extensive infrastructure of IoMT systems requires IDS that can monitor the whole network. \textit{While various types of IDS exist, our method generically applies to all of them. To prevent any confusion, it is important to clarify that the main focus of this paper is on feature engineering techniques for IDS in IoT systems. The intention is not to introduce a new IDS for IoT but to propose a generic feature engineering approach that can be applied effectively in IoT environments.}

Different prior works have shown the importance of feature engineering in improving the IDS’s performance \cite{Hakim} in the context of IIoT security, such as supervisory control and data acquisition (SCADA) systems \cite{Upadhyay} and cloud security \cite{Mishra}. According to Hakim et al. \cite{Hakim}, feature engineering has improved some of the tested models’ accuracies from 51\% to 97\%. Also, the required training time in all models has been almost reduced by half. Thus, developing a feature reduction or feature extraction approach to enhance ML models’ performance is commonly recommended \cite{Parlar}.

Improving IDS's performance can be achieved by using one or more feature reduction methods (i.e., the hybrid technique introduced in Subsection \ref{section:bg}.A), as discussed in \cite{Ravindranath, Padmashree22, Kamarudin, Pawar}. Ravindranath et al. \cite{Ravindranath} propose a feature reduction method that utilizes the whale Pearson hybrid wrapper. This method is based on the binary Whale optimization algorithm, a swarm intelligence algorithm. It reduces the data features from 42 features in the HackerEarth network attack prediction dataset \cite{HackerEarth20} to only 8, with an 8\% accuracy improvement compared to the original dataset using the k-nearest neighbors algorithm. Padmashree and Krishnamoorthi \cite{Padmashree22} propose a decision tree-based Pearson correlation recursive feature elimination model to select a subset of the features to detect various attacks via an optimized deep neural network (DNN) model using the BoT-IoT dataset. This model reduces the number of features in the BoT-IoT dataset to only nine features with 99.20\% accuracy.

Kamarudin et al. \cite{Kamarudin} combine filter and wrapper methods as a single hybrid feature reduction method. This method reduced the number of features from 41 and 33 to 12 and 5 for the KDD CUP’99 and DARPA 1999 datasets, respectively. Also, it enhanced the IDS performance by 9\%.

Another feature reduction method for IDS developed by Pawar et al. \cite{Pawar} selects a subset of features based on a voting scheme from a list of feature selection methods. This scheme reduced the number of features from 41 to 14 for the NSL-KDD dataset and 47 to 18 for the UNSW-NB15 dataset. Yet, none of these feature engineering methods are designed to work on IoT systems.

Another way to design a hybrid feature reduction method is by combining a feature reduction method with some specific artificial neural network (ANN) algorithm. Jingyi et al. \cite{Jingyi} implement a method based on supervised locality-preserving projections and use a backpropagation neural network called an extreme learning machine. Madanan et al. \cite{Madanan}, Abdul Lateef et al. \cite{Lateef20}, Fatani et al. \cite{Fatani21}, and Dahou et al. \cite{Dahou22}  also design similar methods using intelligent water drops, crow swarm optimization algorithms, Aquila optimizer, and reptile search algorithm, respectively. While these methods use the KDD CUP’99 dataset, the Fatani et al. work included three other datasets, including NSL-KDD, BoT-IoT, and CIC2017. Using the KDD CUP’99 dataset, the Fatani et al. method performed the best with an accuracy score of 99.92\% compared to 99.56\%, 92.34\%, 98.58\%, and 98.34\% for Madanan et al., Dahou et al., Jingyi et al., and Abdul Lateef et al., respectively. However, these methods must work with ANN models, which require significant computing power and only work on powerful devices.

Hybrid feature reduction methods are also used to improve the detection rate for medical diagnostics, such as \cite{Shaban,Li20}. Shaban et al. \cite{Shaban}, similar to \cite{Kamarudin}, employ filter and wrapper methods to improve the performance of a KNN model, which is used as a new COVID-19 detection strategy. On the other hand, Li et al. \cite{Li20} illustrate that using multiple feature reduction methods, including PCA in a support vector machine model, can enhance the detection rate for sleep apnea. Nimbalkar et al. \cite{Nimbalkar21} propose a hybrid feature selection method based on the information gain and gain ratio methods to detect DoS and DDoS attacks in IoT systems using the BoT-IoT and KDD Cup 1999 datasets.

In general, our method stands out from other approaches as it significantly improves the performance of IDS in IoT systems. Also, It takes the benefits of both feature selection and extraction methods and reduces their drawbacks. LEMDA is based on embedded methods and achieves a better tradeoff between performance and speed. It offers support for various types of attacks without needing a specific ML model or a complex ANN model. Additionally, it often leads to faster IDS models compared to alternative methods.

%-------------------------------------------------------------------------------
\section{Our Proposed Method}
%-------------------------------------------------------------------------------

Designing a new feature reduction method is essential to enhance the prediction for IDS, especially for IoT systems, since they require fast detection. This makes high accuracy and fast execution indispensable for IDS models.

Our method, LEMDA, is based on MDA and consists of two techniques to satisfy the high accuracy and fast speed requirements of IoT-oriented IDS. The main technique is WEDF, which runs after MDA, where the list of the most informative features is selected. The second one, SF, is an add-on technique to handle the datasets with a categorical feature as the most important feature for the cases when there are a majority of passive attacks like sniffing. In this section, we explain in detail these two techniques. For the rest of the paper, we will use $f_m$ to represent the most informative feature in the list, which is selected by the MDA method, and $f_{mn}$ to represent a new feature, which is created by the WEDF method.

\subsection{Weighted Exponential Decay Formula (WEDF)}
%-------------------------------------

WEDF creates a new feature  $f_{mn}$ based on a predefined dictionary ($WEDF$). This dictionary is constructed from $f_m$ by transforming its samples' values into weights using the exponential decay formula (Eq. \ref{eqn:5}).

\begin{equation}
\label{eqn:5}
 f(x) = ab^x
 \end{equation}

Here, $f(x)$ is the output value (after the decay) in the exponential decay formula, $a$ is the initial value (before the decay), $b$ is the decay factor (a static fraction, $0 < b < 1$, that needs to be set before running the WEDF method e.g., $b=0.5$), and $x$ is the time period (during which $a$ has been decayed). Since $a$ is a static parameter, it can be removed in the WEDF method (i.e., considering $a=1$).

\begin{equation}
\label{eqn:6}
 WEDF_u = f(p)w_u = b^pw_{u} \quad \textnormal{where} \quad {w_{u}}=\frac{z_{u}}{n_{u}}
 \end{equation}
  
Eq. 6 calculates $WEDF_u$, the WEDF score for each $u$ that constitutes $WEDF$. $u$ is a specific unique value from all data instances of the $f_m$ feature. Each $u$ corresponds to a unique data value. In the context of the WUSTL-EHMS dataset, for instance, $u$ can be “TCP,” which is a value of the $f_m$ feature. A more detailed example is provided in the next paragraph. We add a new weight parameter $w_u$ for each unique value $u$ in $f_m$. $z_u$ represents the number of attack samples in the training dataset for each $u$ in $f_m$, and $n_u$ is the total number of samples for each $u$ in $f_m$. Hence, $z_u$ divided by $n_u$ will result in $w_u$ for each $u$. All the weights are sorted in descending order based on $n_u$. Let us denote the index of each $u$ as $p$ (i.e., $p$ ranges from 1 to the number of unique values in $f_m$).

For instance, let us assume that 100 out of 1000 samples in the training dataset have transmission control protocol (TCP) $u~=~TCP$ as their unique value, 10 of which are attack samples; then $z_{TCP}~=~10$, $n_{TCP}~=~100$, and $w_{TCP}~=10/100=0.1$, which will be stored in the $w$ dictionary. Then, assuming TCP is the first unique value in the $w$ (i.e., $p~=~1$) and setting $b~=~0.5$, we can calculate $WEDF_{TCP}$ by $b^pw_{TCP}~=~0.5^1\times0.1~=~0.05$. Hence, all the samples with $u=TCP$ will have $WEDF_{TCP}~=~0.05$. Other unique values in $f_m$ with zero attack samples will have a WEDF score of zero in $f_{mn}$.
 
Finally, the WEDF scores for the $u$ values in the $f_m$ feature using the training dataset are stored in a dictionary ($WEDF$). This dictionary is used to create the $f_{mn}$ feature in the training and testing datasets. Then, the $f_m$ feature is deleted from both datasets. Algorithm \ref{algo:wedf} shows the step-by-step implementation of generating the dictionary.
 
Upon generating the dictionary (assigning a weight to each $u$ based on the proportion of attack samples associated with each value, i.e., $w_u$ to attack samples and 0 to normal samples), the feature distribution for the $f_{mn}$ feature will become roughly a bimodal distribution. Consider an example of a $f_m$ with a standard normal distribution $N(0,1)$. After applying WEDF, the normal samples will tend to cluster around the left side (0.1\% region) of the distribution, and the attack samples will cluster around the right side of the distribution, resulting in a gap in the distribution between the normal and attack samples. This helps the ML model to easily separate the normal samples from the attack samples and increases the importance of the $f_{mn}$ feature compared to the original $f_m$ feature, resulting in better IDS performance.

\begin{algorithm}
\caption{WEDF Method}
\label{algo:wedf}
\begin{algorithmic}[1]
\STATE \textbf{Input:} $f_m$ feature from the Training Dataset ($D_T$)
\STATE \textbf{Output:} $WEDF$ for all $WEDF_u$ in $f_m$ to create $f_{mn}$
\STATE $WEDF, z, w, w_{attack}, w_{normal} = \{\}, \{\}, \{\}, \{\}, \{\}$
\STATE $n \leftarrow value\_counts(D_T[f_m])$ 
%\STATE $z \leftarrow value\_counts(D_T[f_m], label=attack)$
\FOR {$u$ in $n$}
\STATE $z[u] \leftarrow length(D_T[f_m][u], label=attack)$
\IF{$z_u >= 1$}
\STATE $w_{attack}[u] \leftarrow z[u]/n[u]$
\ELSE
\STATE $w_{normal}[u] \leftarrow 0$
\ENDIF
\ENDFOR
\STATE $w \leftarrow concatenate(w_{attack},w_{normal})$
\STATE $p = 1$
\FOR {$u$ in $w$}
\STATE $WEDF[u] \leftarrow b^{p} \times w[u]$ 
\STATE $p = p + 1$
\ENDFOR
\end{algorithmic}
\end{algorithm}

\subsection{Sensitivity Factor (SF)}
%------------------------------------

SF has been added as an add-on besides using the WEDF technique in case the $f_m$ feature is a categorial feature like the \textit{Flags} feature in networking, and most attacks are passive attacks like sniffing. This add-on requires the training and testing datasets, individually, to be arranged in a sequential order, typically based on the timestamps associated with the samples.

SF is also based on the exponential decay formula but without multiplying weights $w$. As shown in Eq. \ref{eqn:7}, we use $d$, an index of the current sample ($s$) based on the last seen suspicious sample, as the input to the exponential decay formula (Eq. \ref{eqn:5}). Similar to the WEDF method, a new feature $f_{smn}$ is created in the training and testing datasets using the $f_{m}$ feature. Using the \textit{Flags} feature in networking as an example, any item in \textit{Flags} that differs from the common values of \textit{Flags}, such as duplicate MACs (M), is considered a suspicious sample. This add-on is used in one of the three datasets, and its results are promising, as shown in the results section.

\begin{equation}
\label{eqn:7}
 SF_s = b^d
 \end{equation}
 
In the case of suspicious samples, the SF score reaches its peak (i.e., a higher SF score indicates a greater likelihood of being an attack sample). Then, the score exponentially decreases for each sample after the suspicious sample(s) until the score reaches zero, as shown in Algorithm \ref{algo:sf}. This is because cyber attacks usually exhibit intensive behaviors over a continuous time period, and the network traffic returns to a normal state after a certain duration.

\begin{algorithm}
\caption{SF Method}
\label{algo:sf}
\begin{algorithmic}[1]
\STATE \textbf{Input:} $f_m$ feature from the Training Dataset ($D_T$)
\STATE \textbf{Output:} $f_{smn}$ feature
\STATE $common\_value \leftarrow$ the most common value in the normal samples
\STATE $b, d = 0, 1$
\FOR {$s$ in $f_m$}
\IF{$s$ is not a $common\_value$ in $f_m$}
\STATE $b, d = 0.5, 1$
\ENDIF
\STATE $s_{new} \leftarrow b^{d}$
\STATE $d= d+1$
\ENDFOR
\end{algorithmic}
\end{algorithm}

%-------------------------------------------------------------------------------
\section{Experimental Methodology}
%-------------------------------------------------------------------------------

In this section, we will demonstrate the evaluation results of the proposed method using three datasets and three models. The datasets are WUSTL-EHMS \cite{Hady}, MQTT-IoT \cite{Hindy}, and BOT-IoT \cite{Koroniotis}. Two of them are collected from general IoT systems (MQTT-IOT and BOT-IoT) and one from an IoMT system (WUSTL-EHMS). Our ML models include decision trees (DT), RF, and two ANN models. We use the $F_1$ score, the Safety score \cite{safety-score}, and the accuracy metrics to compare the performance of feature engineering methods applied to these models.

Using these datasets and models, we evaluate our methods by comparing them to two widely recognized feature reduction techniques, namely PCA and MDA, as well as the scenario where no feature reduction (Base) is applied.

\subsection{Datasets}
%------------------------------------

The three IoT datasets used in our experiments have different sizes starting from 16K to 10M samples and similar numbers of features, as shown in Table \ref{Datasets_Stat}:
\begin{table}[ht]
\caption{Datasets Statistics}
\label{Datasets_Stat}
\begin{center}
\begin{tabular}{ l|l|l } 
 \hline
 Dataset & No. of Samples &  No. of Features\\ 
 \hline
 WUSTL-EHMS & \quad\,\,\,\;16,317 & 44 \\ 
 MQTT-IoT & \,\,\,2,000,000 & 31 \\ 
 BOT-IoT & 10,000,000 & 35 \\ 
 \hline
\end{tabular}
\end{center}
\end{table}
\\
\textbf{1. WUSTL-EHMS:}

This dataset was collected from a real-time enhanced healthcare monitoring system (EHMS) testbed at Washington University in St. Louis, representing a real IDS for the IoMT systems \cite{Hady}. The types of attacks in this dataset are based on man-in-the-middle (MiTM) attacks, such as sniffing and injection attacks. Hence, this dataset has both passive (sniffing only) and active (injection) attacks. This dataset is explained in \cite{Hady} and \cite{WUSTL-EHMS}.
\\
\textbf{2. MQTT-IoT:}

This dataset uses message queuing telemetry transport (MQTT) protocol for IoT systems \cite{Hindy}. The types of attacks in this dataset are as follows: user datagram protocol (UDP) scan, aggressive scan, MQTT brute-force Sparta, and secure shell protocol (SSH) brute-force. The number of samples in this dataset is 20M, but we have randomly selected 2M samples containing all attacks in the original dataset. More about this dataset is available in \cite{Hindy,MQTT-IoT}.
\\
\textbf{3. BOT-IoT:}

This well-known dataset was created using IDS for IoT systems in the Cyber Range Lab of UNSW Canberra. It has different types of attacks, including theft, reconnaissance, denial of service (DoS), and distributed DoS (DDoS) attacks. 10M samples out of 73M samples have been selected to test our method. More about this dataset is available in \cite{Koroniotis,BOT-IoT19}.

The number of selected features for each feature reduction method using these datasets is presented in Table \ref{num_features_datasets}. The Base method represents the method where we use all the features (without feature engineering). It is worth noting that we have removed the identifier features, such as IP addresses and port numbers, from all the methods, including the Base method.

\begin{table}[ht]
\caption{Number of Selected Features per Reduction Method}
\label{num_features_datasets}
\begin{center}
\begin{tabular}{ l|l|l|l|l } 
 \hline
 Dataset & Base & PCA\textsuperscript{*} & MDA & \textbf{LEMDA}\\ 
 \hline
 WUSTL-EHMS & 35 & 14 & 5 & 5+1\textsuperscript{**} \\ 
 MQTT-IoT & 25 & \,\,\,9 & 5 & 5 \\ 
 BOT-IoT & 23 & 10 & 5 & 5 \\ 
 \hline
\end{tabular}
\end{center}
\scriptsize{\qquad\qquad\textsuperscript{*} Explained variance = 95\%}

\scriptsize{\qquad\qquad\textsuperscript{**} Additional feature using the SF method}
%{\raggedright \scriptsize{\qquad\qquad\textsuperscript{*} Explained variance = 95\%}
\end{table}

\subsection{Models}
%----------------------------

We have used DT, RF, and two ANN models. The scikit-learn package has been utilized for DT and RF models with default hyperparameters \cite{sklearnDT}\cite{sklearnRF}. A simple multi-layer perceptron (MLP) model with two layers is used as a simple ANN model. To show the effect of complex ANN models, we have added a Convolutional Neural Network (CNN) model with five layers and Max pooling layers. The Keras package has been utilized for the two ANN models \cite{kerasANN}. More about the hyperparameters are shown in Table \ref{ann_param}.

\textit{As our objective is not to develop optimal machine learning models with optimized hyperparameters but to demonstrate our method's robustness and maintain consistency in experimental comparisons with other methods, we use these simple models with identical parameters across all three datasets.}

\begin{table}[ht]
\caption{ANN Models Hyperparameters}
\label{ann_param}
\begin{center}
\begin{tabular}{ l|l|l } 
 \hline
 Parameter & MLP Typical Value(s) & CNN Typical Value(s)\\ 
 \hline
 \# of layers & 2 & 5 \\ 
 \# of neurons per layer & 20, 1 & 32, 128, 512, 1024, 1\\ 
 \# of epochs & 20 & 2\\
 Kernel size & None & 2\\
 Pooling & None & Max Pooling (size=2)\\
 Batch size & \multicolumn{2}{c}{1000}\\
 Loss function &  \multicolumn{2}{c}{\textit{binary cross-entropy}}\\ 
 Optimizer &  \multicolumn{2}{c}{\textit{Adam}}\\
 Activation function &   \multicolumn{2}{c}{\textit{tanh}, \textit{sigmoid}}\\
\hline
\end{tabular}
\end{center}
\end{table}

The k-fold cross-validation method with ten folds has been utilized on all three models to analyze the models’ performance. In each fold, the dataset is divided into ten subsets; nine of them are used for training and one for testing.

\subsection{Metrics}
We evaluate the models' performances using the accuracy, $F_1$ score, and Safety score. All the metrics are calculated based on the following four categories: true negative (TN), true positive (TP), false positive (FP), and false negative (FN). Label attack is defined as positive, and label normal is defined as negative. TN and TP are the cases when an IDS model correctly predicts a normal sample as normal and an attack sample as an attack, respectively. FP is when the model mistakenly predicts a normal sample as an attack, while FN is when an attack sample is predicted as normal.
\\
\textbf{1. Accuracy:}

This metric represents the ratio of correct predictions to the total predictions, as illustrated in Eq. \ref{eqn:8}.
\begin{equation}
\label{eqn:8}
\mbox{Accuracy}=\frac{\mbox{TN}+\mbox{TP}}{\mbox{FP}+\mbox{FN}+\mbox{TN}+\mbox{TP}} 
 \end{equation} 
\\
\textbf{2. $F_1$ score:}

This metric is popular for security applications such as IDS. It is the harmonic mean between recall and precision, as illustrated in Eq. \ref{eqn:9}.
\begin{equation}
\label{eqn:9}
F_1\mbox{ score}=2\times \frac{\mbox{Precision}\times \mbox{Recall}}{\mbox{Precision}+\mbox{Recall}}=\frac{\mbox{TP}}{\mbox{TP}+\frac{1}{2}\mbox{(FP+FN)}}
 \end{equation} 
\\
\textbf{3. Safety score:}

The safety score is designed by Salman et al. \cite{safety-score} specifically for security applications to fulfill the shortcoming of other metrics in these types of applications. This metric adds weights (Eq. \ref{eqn:10}) to the primary four model's outcome categories (TP, TN, FP, and FN). \begin{equation}
\label{eqn:10}
\mbox{Safety score}=\frac{w_{\mbox{\tiny TN}}\mbox{TN}+w_{\mbox{\tiny TP}}\mbox{TP}}{w_{\mbox{\tiny FP}}\mbox{FP}+w_{\mbox{\tiny FN}}\mbox{FN}+w_{\mbox{\tiny TN}}\mbox{TN}+w_{\mbox{\tiny TP}}\mbox{TP}} 
 \end{equation} 

For generality, we assign the following weights assuming that both FN and FP have the same importance, as explained in \cite{Hady} and \cite{safety-score}: 
\[w_{\mbox{\tiny TN}}=\frac{1}{19}, w_{\mbox{\tiny TP}}=\frac{2}{19}, w_{\mbox{\tiny FP}}=\frac{8}{19}, w_{\mbox{\tiny FN}}=\frac{8}{19}\]

\begin{table}[!ht]
\caption{List of Selected Features}
\label{listselectedfeatures}
\begin{center}
\begin{tabular}{ l|l|l|l} 
\hline
     & \multicolumn{3}{c}{Dataset}\\
    \#&WUSTL-EHMS&MQTT-IoT&BOT-IoT\\\hline
    \textbf{1}&\textit{\textbf{Flgs}} & \textit{\textbf{protocol}} & \textit{\textbf{state}}
    \\2&\textit{DIntPkt} & \textit{mqtt\_messagetype} & \textit{sbytes}
    \\3&\textit{DstJitter} & \textit{ttl} & \textit{bytes}
    \\4&\textit{Rate} & \textit{mqtt\_messagelength} & \textit{proto}
    \\5&\textit{DstLoad} & \textit{ip\_len} & \textit{srate}\\
 \hline
\end{tabular}
\end{center}
\end{table}

\begin{table*}[!hb]
\caption{Methods Results For the Three Datasets}
\label{methods_results}
\begin{center}
\begin{tabular}{ l|l|l|l|l|l|l|l|l|l|l }
\hline
\multicolumn{11}{c}{WUSTL-EHMS Dataset}\\
 \hline
Model & Method & TP &  FN &  FP & TN & Accuracy & $F_1$ & Safety  & Training Time & Detection Time\\ 
 \hline
 \multirow{4}{*}{DT}& Base & \quad\,\;127 & \quad\,\;\,\;78 & \,\;\,\;358 &\,\;1069 & 73.284\% & 36.812\% & 27.499\% & \quad0.247 & \,\;0.000219\\
    &PCA\textsuperscript{*} & \quad\,\;109 & \quad\,\;\,\;95 & \,\;\,\;244 & \,\;1184 & 79.228\% & 39.138\% & 34.079\% & \quad0.336 & \,\;0.000173\\
			      &MDA & \quad\,\;125 & \quad\,\;\,\;80 & \,\;\,\;288 & \,\;1139 & 77.451\% & 40.453\% & 32.056\% & \quad0.082 & \,\;0.000194\\
	      &\textbf{LEMDA}& \quad\,\;173 & \quad\,\;\,\;32 & \,\;\,\;43 & \,\;1384 & \textbf{95.404\%} & \textbf{82.185\%} & \textbf{74.249\%} & \textbf{\quad0.069} & \textbf{\,\;0.000132}\\
 \hline
 \multirow{4}{*}{RF}& Base & \quad\,\;115 & \quad\,\;\,\;90 & \,\;\,\;125 & \,\;1302 & 86.826\% & 51.685\% & 47.109\% & \quad0.451 & \,\;0.020998\\
    &PCA\textsuperscript{*} & \quad\,\;105 & \quad\,\;100 & \,\;\,\;139 & \,\;1288 & 85.355\% & 46.771\% & 43.930\% & \quad0.661 & \,\;0.020724\\
			      &MDA & \quad\,\;122 & \quad\,\;\,\;83 & \,\;\,\;189 &\,\;1238 & 83.333\% & 47.287\% & 40.514\% & \quad0.386 & \,\;0.021060\\
	     &\textbf{LEMDA}& \quad\,\;183 & \quad\,\;\,\;22 & \quad\,\;37 & \,\;1390 & \textbf{96.385\%} & \textbf{86.118\%} & \textbf{78.815\%} & \textbf{\quad0.287} & \textbf{\,\;0.020516}\\
 \hline
  \multirow{4}{*}{MLP}& Base & \quad\,\;104 & \quad\,\;101 & \,\;\,\;179 & \,\;1248 & 82.843\% & 42.623\% & 39.394\% & \quad1.716 & \,\;0.067576\\
        &PCA\textsuperscript{*} & \quad\,\;105 & \quad\,\;100 & \,\;\,\;221 & \,\;1206 & 80.331\% & 39.548\% & 35.542\% & \quad1.238 & \,\;0.071022\\
			     &MDA & \quad\,\;150 & \quad\,\;\,\;55 & \,\;\,\;730 & \,\;\,\;697 & 51.900\% & 27.650\% & 13.701\% & \quad1.229 & \textbf{\,\;0.063846}\\
	  &\textbf{LEMDA}& \quad\,\;116 & \quad\,\;\,\;89 & \quad\,\;15 & \,\;1412 & \textbf{93.627\%} & \textbf{69.048\%} & \textbf{66.397\%} & \textbf{\quad1.213} & \,\;0.064459\\
\hline
  \multirow{4}{*}{CNN}& Base & \quad\,\;134 & \quad\,\;\,\;71 & \,\;\,\;546 & \,\;\,\;881 & 62.194\% & 30.282\% & 18.882\% & \quad2.369 & \,\;0.142412\\
        &PCA\textsuperscript{*} & \quad\,\;159 & \quad\,\;\,\;46 & \,\;1018 & \,\;\,\;409 & 34.804\% & 23.010\% & \,\;7.869\% & \quad1.498 & \,\;0.128004\\
			     &MDA & \quad\,\;180 & \quad\,\;\,\;25 & \,\;1142 & \,\;\,\;285 & 28.493\% & 23.576\% & \,\;6.462\% & \quad1.067 & \textbf{\,\;0.106574}\\
	  &\textbf{LEMDA}& \quad\,\;126 & \quad\,\;\,\;79 & \quad\quad7 & \,\;1420 & \textbf{94.730\%} & \textbf{74.556\%} & \textbf{70.847\%} & \textbf{\quad1.068} & \,\;0.120796\\
 \hline
 \hline
\multicolumn{11}{c}{MQTT-IoT Dataset}\\
 \hline
 Model & Method & TP &  FN &  FP & TN & Accuracy & $F_1$ & Safety  & Training Time & Detection Time\\ 
 \hline
 \multirow{4}{*}{DT}& Base & \,\;19596 & 160651 & \,\;2383 & 17370 & 18.483\% & 19.380\% & \,\;4.156\% & \quad4.251 & \,\;0.008660\\
    &PCA\textsuperscript{*} & 113441 & \,\;66806 & \,\;6508 & 13245 & 63.343\% & 75.578\% & 29.049\% & \quad3.221 & \,\;0.006749\\
			      &MDA & \,\;20361 & 159886 & \,\;2463 & 17290 & 18.826\% & 20.053\% & \,\;4.276\% & \textbf{\quad1.038} & \,\;0.004655\\
&\textbf{LEMDA}& 160890 & \,\;19357 & \,\;9984 & \,\;9769 & \textbf{85.330\%} & \textbf{91.644\%} & \textbf{58.549\%} & \quad1.077 & \textbf{\,\;0.004508}\\
 \hline
 \multirow{4}{*}{RF}& Base & \,\;55720 & 124527 & \,\;2061 & 17692 & 36.706\% & 46.818\% & 11.309\% & \,\;58.882 & \,\;0.124873\\
    &PCA\textsuperscript{*} & \,\;57690 & 122558 & \,\;2093 & 17659& 37.675\% & 48.069\% & 11.771\% & \,\;60.652 & \,\;0.119515\\
			      &MDA & \,\;20234 & 160013 & \,\;2433 & 17320 & 18.777\% & 19.943\% & \,\;4.257\% & \,\;31.612 & \,\;0.109121\\
&\textbf{LEMDA}& 178417 & \,\;\,\;1830 & 12052 & \,\;7701 & \textbf{93.059\%} & \textbf{96.255\%} & \textbf{76.649\%} & \textbf{\,\;24.630} & \textbf{\,\;0.107833}\\
 \hline
  \multirow{4}{*}{MLP}& Base & \,\;64395 & 115852 & \,\;1088 & 18665 & 41.530\% & 52.411\% & 13.616\% & \,\;40.425 & \,\;1.752910\\
     &PCA\textsuperscript{*} & 112778 & \,\;67469 & \,\;4794 & 14959 & 63.869\% & 75.736\% & 29.381\% & \,\;38.434 & \,\;1.727500\\
			      &MDA & \,\;16252 & 163995 & \,\;\,\;629 & 19124 & 17.688\% & 16.489\% & \,\;3.772\% & \textbf{\,\;37.485} & \,\;1.737760\\
&\textbf{LEMDA}& 178972 & \,\;\,\;1276 & 12061 & \,\;7691 & \textbf{93.332\%} & \textbf{96.408\%} & \textbf{77.411\%} & \,\;37.505 & \textbf{\,\;1.708496}\\
\hline
  \multirow{4}{*}{CNN}& Base & 115844 & \,\;64404 & \,\;4715 & 15037 & 65.441\% & 77.022\% & 30.853\% & 181.872 & \,\;6.220485\\
     &PCA\textsuperscript{*} & \,\;94843 & \,\;85404 & \,\;8631 & 11122 & 52.983\% & 66.857\% & 21.069\% & 108.087 & \,\;5.084061\\
			      &MDA & \,\;72885 & 107362 & \,\;4843 & 14910 & 43.898\% & 56.505\% & 15.183\% & \textbf{\,\;72.776} & \textbf{\,\;4.669276}\\
&\textbf{LEMDA}& 179002 & \,\;\,\;1246 & 12071 & \,\;7681 & \textbf{93.341\%} & \textbf{96.414\%} & \textbf{77.439\%} & \,\;72.949 & \,\;4.673046\\
 \hline \hline

\multicolumn{11}{c}{BOT-IoT Dataset}\\
 \hline
Model & Method & TP &  FN &  FP & TN & Accuracy & $F_1$ & Safety & Training Time & Detection Time\\ 
 \hline
 \multirow{4}{*}{DT}& Base & 917734 & \,\;82136 & \quad\quad5 &\,\;\,\;125 & 91.786\% & 95.716\% & 73.638\% & 175.926 & \,\;0.050922\\
  &PCA\textsuperscript{*} & 965000 &  \,\;34870 &  \quad\,\;42 & \quad\,\;88 & 96.509\% & 98.223\% & 87.359\% & 392.076 & \,\;0.059293\\
			      &MDA & 950423 &  \,\;49447 &  \quad\quad5 & \,\;\,\;125 & 95.055\% & 97.464\% & 82.774\% & \;\;38.299 & \,\;0.038469\\
	       &\textbf{LEMDA}& 999854 &  \quad\,\;\,\;16 &  \quad\quad7 & \,\;\,\;123 & \textbf{99.998\%} & \textbf{99.999\%} & \textbf{99.991\%} & \;\;\textbf{36.079} & \textbf{\,\;0.036017}\\
 \hline
 \multirow{4}{*}{RF}& Base & 981079 & \,\;18791 & \quad\quad3 &\,\;\,\;127 & 98.121\% & 99.051\% & 92.883\% & 445.560 & \,\;0.713076\\
 &PCA\textsuperscript{*} & 984417 &  \,\;15453 &  \quad\,\;42 & \quad\,\;88 & 98.451\% & 99.219\% & 94.077\% & 982.130 & \,\;0.829192\\
			      &MDA & 962375 &  \,\;37495 &  \quad\quad2 & \,\;\,\;128 & 96.250\% & 98.089\% & 86.517\% & 311.985 & \,\;0.635074\\
	       &\textbf{LEMDA}& 999857 &  \quad\,\;\,\;13 &  \quad\,\;11 & \,\;\,\;119 & \textbf{99.998\%} & \textbf{99.999\%} & \textbf{99.990\%} & \textbf{152.270} & \textbf{\,\;0.611572}\\
 \hline
  \multirow{4}{*}{MLP}& Base & 938838 & \,\;61032 & \quad\quad7 &\,\;\,\;123 & 93.896\% & 96.852\% & 79.362\% & 223.130 & \,\;8.510574\\
 	&PCA\textsuperscript{*} & 936317 &  \,\;63553 &  \quad\,\;28 & \,\;\,\;102 & 93.642\% & 96.716\% & 78.641\% & \textbf{202.012} & \,\;8.656834\\
			      &MDA & 499925 &  499945 &  \quad\,\;22 & \,\;\,\;108 & 50.003\% & 66.665\% & 20.000\% & 206.871 & \,\;8.560095\\
	       &\textbf{LEMDA}& 998828 &  \,\;\,\;1042 &  \quad\,\;47 & \quad\,\;83 & \textbf{99.891\%} & \textbf{99.946\%} & \textbf{99.566\%} & 207.164 & \textbf{\,\;8.422081}\\
\hline
  \multirow{4}{*}{CNN}& Base & 964703 & \,\;35167 & \quad\,\;26 &\,\;\,\;104 & 96.481\% & 98.209\% & 87.267\% & 964.367 & 41.930902\\
 	&PCA\textsuperscript{*} & 930038 &  \,\;69832 &  \quad\,\;27 & \,\;\,\;103 & 93.014\% & 96.380\% & 76.897\% &619.433 & 25.352587\\
			      &MDA & 487379 &  512491 &  \quad\,\;10 & \,\;\,\;120 & 48.750\% & 65.541\% & 19.210\% & 410.698 & 23.708180\\
	       &\textbf{LEMDA}& 977230 &  \,\;22640 &  \quad\,\;27 & \,\;\,\;103 & \textbf{97.733\%} & \textbf{98.854\%} & \textbf{91.510\%} &  \textbf{403.539} & \textbf{23.344931}\\
 \hline

 \end{tabular}
\end{center}
{\raggedright \scriptsize{\qquad\quad\qquad\textsuperscript{*} Explained variance = 95\%}}
\end{table*}

\section{Experimental Results}
%----------------------------------------
The evaluation results are categorized into five subsections. The first three subsections present the results for each dataset individually, considering all the methods. The fourth subsection focuses on comparing the training and detection times across all datasets and models, employing all the methods. Lastly, the fifth subsection provides a comprehensive comparison with previous works.

For the metrics and time comparisons, we use the average of the k-fold cross-validation with ten folds. We set the $b$ decay factor for WEDF and SF to be 0.5 as a balanced tradeoff between FN and FP values. Further investigation involved conducting multiple trials within the range of 0.1 to 0.9. Through these trials, we found that the FP value increased and the FN value decreases as $b$ increases and vice versa. Since we have set both values to be equally important, as explained in the previous section, the $b$ value is set to be 0.5 for WEDF and SF. For other applications, the $b$ value should be set based on the importance of the FN value over the FP value or vice versa.

The selected features using the MDA method and our method for all three datasets are shown in Table \ref{listselectedfeatures}. Note that our method creates an $f_{mn}$ feature for each dataset using their $f_{m}$ features, which are shown in the first row of Table \ref{listselectedfeatures}.

\subsection{WUSTL-EHMS:}

The WUSTL-EHMS dataset, as shown in Table \ref{listselectedfeatures}, has the \textit{Flgs} feature as the most informative feature $f_{m}$, and the majority of the attacks are passive attacks (sniffing). As a result, among the three datasets, it is the only one that is suitable to use the SF add-on with the WEDF method. Across the three models in Table \ref{methods_results}, our method shows average values for accuracy, $F_1$ score, and Safety score of approximately 95\%, 78\%, and 73\%, respectively. These outcomes show that our method has an average improvement of  about 28\%, 52\%, and 61\% in accuracy, $F_1$ score, and Safety score, respectively, compared to the other methods.

DT, RF, and ANN models show similar performances across the Base, PCA, and MDA methods, while our method outperforms all of them. As seen in Table \ref{methods_results}, our method performs almost twice better with the security-oriented metric Safety score than the other methods in the three models. With the accuracy and the $F_1$ score, our method still significantly shows improved results compared to other methods.

\subsection{MQTT-IoT:}

Similar to the WUSTL-EHMS dataset results, across the three models, our method showcases average accuracy, $F_1$ score, and Safety score of approximately 91\%, 95\%, and 73\% across the three models. MQTT-IoT results showed that our method has outperformed other methods by at least 50\% using the $F_1$ score, as illustrated in Table \ref{methods_results}. Furthermore, even when the MDA method uses the same 4 out of 5 features as our method, the difference in performance between them reached almost 70\% on average. The average improvements of our method using the accuracy and Safety score are 56\% and 79\%, respectively.

\subsection{BOT-IoT:}

This dataset has more attacks, such as DoS and DDoS attacks, making it more general for IoT systems than the other two datasets. Our method demonstrates an average performance of approximately 99\% for accuracy, 99\% for the $F_1$ score, and 98\% for the Safety score across the three models. Even here, our method shows clear performance improvement in the DT and RF models compared to the other methods in terms of accuracy, $F_1$ score, and Safety score, as shown in Table \ref{methods_results}.

Given these results and the varying attacks in each dataset, our method demonstrates superior performance, rendering it more suitable as an IDS for IoT systems using AI-based models compared to other methods.

In particular, by comparing the results of the MLP and CNN models across all three datasets, the CNN exhibits superior performance over the MLP in the two large datasets, MQTT-IoT and Bot-IoT. This suggests that for large datasets without feature engineering, a more complex ANN model is beneficial. Additionally, LEMDA exhibits enhancement (from MLP to CNN) in two datasets, WUSTL-EHMS and MQTT-IoT. However, this increase in complexity leads to very high training and detection times. 

These findings confirm that feature engineering methods are essential to reduce the computational complexity with simpler models. While more complex models may not necessarily enhance performance, they still contribute to this reduction of computing time.

\subsection{Training and Detection Time Comparison:}

The improvement in model performance is not the only important requirement for IDS in IoT systems since the models’ training and detection time are also critical. As presented in Table \ref{methods_results}, our method achieves the lowest or very close to the lowest training time compared to other methods, with an average of 0.66s, 34.04s, and 199.76s in WUSTL-EHMS, MQTT-IoT, and BOT-IoT datasets, respectively. 

The detection times are also shown in Table \ref{methods_results}. Similar to the average training time, our model detection times are the lowest in almost all the cases, with an average of 0.05s, 1.62s, and 8.10s in WUSTL-EHMS, MQTT-IoT, and BOT-IoT datasets, respectively. This lets us conclude that our method enhances the IDS performance and takes less time to train and detect attacks using different models in most cases. Hence, it makes the IDS models very accurate and fast to train ML models and detect attacks.

\subsection{Related Work Comparison:}

In addition to comparing our work with PCA and MDA, we further assess its performance against four related works \cite{Padmashree22}, \cite{Fatani21},  \cite{Dahou22}, and \cite{Nimbalkar21}, using the BoT-IoT dataset, as presented in Table \ref{related_work_comparison}. As mentioned in Section \ref{section:related_work}, \cite{Fatani21}, \cite{Dahou22}, and \cite{Nimbalkar21} used only one ML classifier model to report their results, while \cite{Padmashree22} method uses DNN classifier. In contrast, we have tested our method with multiple ML models, including DT and RF, in addition to the two ANN models, including MLP and CNN, to show the versatility of our method.

All the attacks were included in \cite{Padmashree22}, \cite{Fatani21}, and \cite{Dahou22} results, while only DoS/DDoS attacks were included in \cite{Nimbalkar21}. The methods proposed by \cite{Padmashree22}, \cite{Fatani21}, and \cite{Dahou22} require using an ANN model as part of the implementation of their methods. The methods in \cite{Fatani21} and \cite{Dahou22} were built to transform the feature space before selecting the best set of features. On the other hand, our method will only transform the most informative feature after the selection process is completed by the MDA method. Compared to these methods, our work shows comparable or better results than these works with up to 85\% feature reduction rate using ML and ANN models.

\begin{table*}[!ht]
\caption{Related Work Comparison using BoT-IoT Dataset}
\label{related_work_comparison}
\begin{center}
\begin{tabular}{ l|l|l|l|l|l }
 \hline
  & LEMDA & \cite{Padmashree22} & \cite{Fatani21} & \cite{Dahou22} & \cite{Nimbalkar21} \\ 
 \hline
 Category & Supervised & Supervised & Supervised & Supervised & Supervised\\
 Type of attacks & All & All & All & All & DoS/DDoS\\
 Model & DT/RF/ANN & DNN & KNN & KNN & JRip\\
 Require ANN & No & Yes & Yes & Yes & No\\
 Reduction Approach & Selection+Extraction &  Only selection &  Selection+Extraction & Selection+Extraction & Only selection\\
Reduction Method &  MDA+WEDF & DT-PCRFE & AQU+CNN &  RSA+CNN & IG+GR\\
No. of Samples & 10M & 13.9M & 3.6M & 3.6M & 0.7M\\
Feature Reduction Rate & 85.7\% & 74.3\% & -----  & -----  & 54.3\%\\
No. of Features & 5 & 9 & ----- & ----- & 16 \\
Accuracy & 99.998\% [DT/RF] & 99.200\% & 99.994\% & 99.993\% & 99.999\%\\ 
$F_1$ & 99.999\% [DT/RF] & 98.910\% & 99.992\% & 99.992\% & ----- \\ 

 \hline
\end{tabular}
\end{center}
\end{table*}

%\clearpage
%-------------------------------------------------------------------------------
\section{Conclusion}
%-------------------------------------------------------------------------------

IDS models for IoT systems require faster training and detecting time along with high performance. Therefore, these require specialized feature reduction methods. This paper presented a new feature reduction method called LEMDA. Our proposed method uses two new techniques called WEDF and SF to generate a representative feature based on the most informative feature from the MDA method. We used three different datasets with different sizes, three different ML models, and three different metrics. We compare our method with other methods, including MDA, PCA, and a base method without feature reduction methods as the ground truth of our experimental results. Our results show that LEMDA performs better than the other methods in all three datasets and ML models by an average of 34\%, 57\%, and 56\% using the $F_1$, the Safety scores, and the accuracy scores, respectively. Furthermore, the proposed method achieved the lowest required training and detection times in most cases, making it run faster than other methods.

For future work, we plan to investigate the improvement of our method using best-optimized models and then compare the results with the plain models. We will examine the potential of our method for semi-supervised and unsupervised ML-based IDS. Moreover, applying our method to applications (other than IoT systems) can help determine its limits.

%\begin{table}[!htbp]

%-------------------------------------------------------------------------------
\section*{Acknowledgments}
%-------------------------------------------------------------------------------
This work was supported in part by the grant ID NPRP13S-0205-200265 funded by the Qatar National Research Fund (QNRF) and by Prince Salam Bin Abdulaziz University, Al-Kharj, Saudi Arabia. The statements made herein are solely the responsibility of the authors.


\begin{thebibliography}{1}
\bibliographystyle{IEEEtran}

\bibitem{IoTAnalytics2022}
E. Schiller, A. Aidoo, J. Fuhrer, J. Stahl, M. Ziörjen, and B. Stiller, “Landscape of IoT security,”
Computer Science Review, vol. 44, 100467, 2022.

%\bibitem{IoTAnalytics2021}
%IoT Analytics, “State of IOT – summer 2021,” IoT Analytics, 01-Nov-2021. [Online]. Available: https://iot-analytics.com/product/state-of-iot-summer-2021/. [Accessed: 02-Sep-2022]. 

\bibitem{Lincbi}
F. Li, R. Xie, Z. Wang, L. Guo, J. Ye, P. Ma, and W. Song, “Online distributed IoT security monitoring with Multidimensional Streaming Big Data,” IEEE Internet of Things Journal, vol. 7, no. 5, pp. 4387–4394, 2020.

\bibitem{Thakkar}
A. Thakkar and R. Lohiya, “A review on machine learning and Deep Learning Perspectives of IDS for IoT: Recent updates, security issues, and challenges,” Archives of Computational Methods in Engineering, vol. 28, no. 4, pp. 3211–3243, 2020. 

\bibitem{Vijayanand}
R. Vijayanand and D. Devaraj, “A Novel Feature Selection Method Using Whale Optimization Algorithm and Genetic Operators for Intrusion Detection System in Wireless Mesh Network,” in IEEE Access, vol. 8, pp. 56847-56854, 2020.

\bibitem{CZ23}
J. Cui, L. Zong, J. Xie, and M. Tang, “A novel multi-module integrated intrusion detection system for high-dimensional imbalanced data,” Applied Intelligence, vol. 53, no. 1, pp. 272-288, 2023/01/01 2023, doi: 10.1007/s10489-022-03361-2.
%\bibitem{Gupta2020}
%A. Gupta, “Feature selection techniques in machine learning,” Analytics Vidhya, 10-Oct-2020. [Online]. Available: https://www.analyticsvidhya.com/blog/2020/10/feature-selection-techniques-in-machine-learning/. [Accessed: 10-Jan-2022]. 

\bibitem{Yang21}
P. Yang, H. Huang, and C. Liu, “Feature selection revisited in the single-cell era,” Genome Biology, vol. 22, no. 1, p. 321, 2021.

\bibitem{Shaban}
W. M. Shaban, A. H. Rabie, A. I. Saleh, and M. A. Abo-Elsoud, “A new COVID-19 patients detection strategy (CPDS) based on hybrid feature selection and Enhanced KNN classifier,” Knowledge-Based Systems, vol. 205, p. 106270, 2020.

\bibitem{Kamarudin}
M. H. Kamarudin, C. Maple, and T. Watson, “Hybrid feature selection technique for intrusion detection system,” International Journal of High-Performance Computing and Networking, vol. 13, no. 2, pp. 232-240, 2019.

\bibitem{Li20}
X. Li, S. H. Ling, and S. Su, “A Hybrid Feature Selection and Extraction Methods for Sleep Apnea Detection Using Bio-Signals,” Sensors, vol. 20, no. 15, p. 4323, Aug. 2020.

\bibitem{Shekhawat21}
S. S. Shekhawat, H. Sharma, S. Kumar, A. Nayyar and B. Qureshi, “bSSA: Binary Salp Swarm Algorithm With Hybrid Data Transformation for Feature Selection,” in IEEE Access, vol. 9, pp. 14867-14882, 2021.

\bibitem{Hady}
A. A. Hady, A. Ghubaish, T. Salman, D. Unal, and R. Jain, “Intrusion Detection System for Healthcare Systems Using Medical and Network Data: A Comparison Study,” in IEEE Access, vol. 8, pp. 106576-106584, 2020.

\bibitem{Farrukh}
Y. A. Farrukh, Z. Ahmad, I. Khan, and R. M. Elavarasan, “A Sequential Supervised Machine Learning Approach for Cyber Attack Detection in a Smart Grid System,” 2021 North American Power Symposium (NAPS), 2021, pp. 1-6.

%\bibitem{WikiDT}
%Wikipedia, “Decision tree learning,” 2021. [Online]. Available: https://en.wikipedia.org/wiki/Decision\_tree\_learning\#\\Gini\_impurity. [Accessed: 14-Jan-2022]. 

%\bibitem{Scheidel}
%C. Scheidel, “Be aware of bias in RF variable importance metrics,” pi: predict/infer, 20-Jun-2018. [Online]. Available: https://blog.methodsconsultants.com/posts/be-aware-of-bias-in-rf-variable-importance-metrics/. [Accessed: 14-Jan-2022]. 

%\bibitem{scikitFE}
%scikit, “Feature importances with a forest of trees,” 2021. [Online]. Available: https://scikit-learn.org/stable/auto\_examples/ensemble/\\plot\_forest\_importances.html. [Accessed: 14-Jan-2022].

\bibitem{Ward21}
I. R. Ward, L. Wang, J. Lu, M. Bennamoun, G. Dwivedi, and F. M. Sanfilippo, “Explainable artificial intelligence for pharmacovigilance: What features are important when predicting adverse outcomes?"
Computer Methods and Programs in Biomedicine, vol 212, 2021, p. 106415.

%\bibitem{Han}
%H. Han, X. Guo, and H. Yu, “Variable selection using mean decrease accuracy and mean decrease Gini based on random forest,” 2016 7th IEEE International Conference on Software Engineering and Service Science (ICSESS), 2016.

%\bibitem{scikitMDA}
%scikit, “4.2. Permutation feature importance,” 2021. [Online]. Available: https://scikit-learn.org/stable/modules/permutation\_importance.html\#\\permutation-importance. [Accessed: 14-Jan-2022]. 

%\bibitem{Kumar20}
%A. Kumar, “PCA explained variance concepts with python example,” Data Analytics, 08-Aug-2020. [Online]. Available: https://vitalflux.com/pca-explained-variance-concept-python-example/. [Accessed: 15-Jan-2022]. 

%\bibitem{Dubey18}
%A. Dubey, “The mathematics behind Principal Component Analysis,” Medium, 25-Dec-2018. [Online]. Available: https://towardsdatascience.com/the-mathematics-behind-principal-component-analysis-fff2d7f4b643. [Accessed: 15-Jan-2022]. 

%\bibitem{Mikulski19}
%B. Mikulski, “PCA-how to choose the number of components?,” Bartosz Mikulski, 03-Jun-2019. [Online]. Available: https://www.mikulskibartosz.name/pca-how-to-choose-the-number-of-components/. [Accessed: 15-Jan-2022]. 

\bibitem{Vilsen21}
S. B. Vilsen and D.-I. Stroe, “Battery state-of-health modelling by multiple linear regression,” Journal of Cleaner Production, vol. 290, p. 125700, 2021.

\bibitem{Swain21}
D. Swain, N. Chillur, S. Patel and A. Bhilare, “Intelligent System for Detecting Intrusion with Feature Bagging,” 2021 International Conference on Artificial Intelligence and Machine Vision (AIMV), 2021, pp. 1-4, doi: 10.1109/AIMV53313.2021.9670940.

\bibitem{Nguyen22}
X.-H. Nguyen, X.-D. Nguyen, H.-H. Huynh, and K.-H. Le, “Realguard: A Lightweight Network Intrusion Detection System for IoT Gateways,” Sensors, vol. 22, no. 2, p. 432, 2022. [Online]. Available: https://www.mdpi.com/1424-8220/22/2/432.

\bibitem{Yu22}
J. Yu, X. Ye, and H. Li, “A high precision intrusion detection system for network security communication based on multi-scale convolutional neural network,” Future Generation Computer Systems, vol. 129, pp. 399-406, 2022/04/01/ 2022, doi: https://doi.org/10.1016/j.future.2021.10.018.

\bibitem{Lightbody22}
D. Lightbody, D. -M. Ngo, A. Temko, C. Murphy and E. Popovici, “Host-Based Intrusion Detection System for IoT using Convolutional Neural Networks,” 2022 33rd Irish Signals and Systems Conference (ISSC), Cork, Ireland, 2022, pp. 1-7, doi: 10.1109/ISSC55427.2022.9826188.

\bibitem{Mandal21}
S. Mandal, A. Sai Sabitha, and D. Mehrotra, “Analysis on Protocol-Based Intrusion Detection System Using Artificial Intelligence,” in Machine Intelligence and Smart Systems, Singapore, S. Agrawal, K. Kumar Gupta, J. H. Chan, J. Agrawal, and M. Gupta, Eds., 2021// 2021: Springer Nature Singapore, pp. 131-143. 

\bibitem{Zeeshan22}
M. Zeeshan et al., “Protocol-Based Deep Intrusion Detection for DoS and DDoS Attacks Using UNSW-NB15 and Bot-IoT Data-Sets,” in IEEE Access, vol. 10, pp. 2269-2283, 2022, doi: 10.1109/ACCESS.2021.3137201.

\bibitem{Hakim}
L. Hakim, R. Fatma, and Novriandi, “Influence Analysis of Feature Selection to Network Intrusion Detection System Performance Using NSL-KDD Dataset,” 2019 International Conference on Computer Science, Information Technology, and Electrical Engineering (ICOMITEE), 2019, pp. 217-220.

\bibitem{Upadhyay}
D. Upadhyay, J. Manero, M. Zaman, and S. Sampalli, “Gradient Boosting Feature Selection With Machine Learning Classifiers for Intrusion Detection on Power Grids,” in IEEE Transactions on Network and Service Management, vol. 18, no. 1, pp. 1104-1116, 2021.

\bibitem{Mishra}
P. Mishra, V. Varadharajan, E. S. Pilli, and U. Tupakula, “VMGuard: A VMI-Based Security Architecture for Intrusion Detection in Cloud Environment,” in IEEE Transactions on Cloud Computing, vol. 8, no. 3, pp. 957-971, 1 July-Sept. 2020.

\bibitem{Parlar}
T. Parlar and E. Sarac, “IWD based feature selection algorithm for sentiment analysis,” Elektronika ir Elektrotechnika, vol. 25, no. 1, 2019.

\bibitem{Ravindranath}
V. Ravindranath, S. Ramasamy, R. Somula, K. S. Sahoo, and A. H. Gandomi, “Swarm Intelligence Based Feature Selection for Intrusion and Detection System in Cloud Infrastructure,” 2020 IEEE Congress on Evolutionary Computation (CEC), 2020, pp. 1-6.

\bibitem{Padmashree22}
A. Padmashree and M. Krishnamoorthi, “Decision Tree with Pearson Correlation-based Recursive Feature Elimination Model for Attack Detection in IoT Environment,” Information Technology and Control, vol. 51, no. 4, pp. 771-785, 2022.

\bibitem{Pawar}
Y. Pawar, N. Zamzami, and N. Bouguila, “An Effective Hybrid Anomaly Detection System Based on Mixture Models,” 2020 International Symposium on Networks, Computers and Communications (ISNCC), 2020, pp. 1-6.

\bibitem{HackerEarth20}
HackerEarth, “Predict Network Attacks,” 2020. [Online]. Available: https://www.hackerearth.com/problem/machine-learning/sample/. [Accessed: 03-Oct-2023]. 

\bibitem{Jingyi}
W. Jingyi, G. Xusheng, H. Jieli, and L. Shenghou, “ELM Network Intrusion Detection Model Based on SLPP Feature Extraction,” 2021 IEEE International Conference on Power, Intelligent Computing and Systems (ICPICS), 2021, pp. 46-49.

\bibitem{Madanan}
M. Madanan, A. Venugopal, and N. C.Velayudhan, “A Hybrid Anomaly Based Intrusion Detection Methodology Using IWD for LSTM Classification,” 2020 IEEE International Conference on Advanced Networks and Telecommunications Systems (ANTS), 2020, pp. 1-5.

\bibitem{Lateef20}
A. A. Abdul Lateef, S. T. Faraj Al-Janabi, and B. Al-Khateeb, “Hybrid Intrusion Detection System Based on Deep Learning,” 2020 International Conference on Data Analytics for Business and Industry: Way Towards a Sustainable Economy (ICDABI), 2020, pp. 1-5.

\bibitem{Fatani21}
A. Fatani, A. Dahou, M. A. A. Al-qaness, S. Lu, and M. A. Abd Elaziz, “Advanced Feature Extraction and Selection Approach Using Deep Learning and Aquila Optimizer for IoT Intrusion Detection System,” Sensors, vol. 22, no. 1, p. 140, Dec. 2021, doi: 10.3390/s22010140.

\bibitem{Dahou22}
A. Dahou et al., “Intrusion Detection System for IoT Based on Deep Learning and Modified Reptile Search Algorithm,” Computational Intelligence and Neuroscience, vol. 2022, p. 6473507, 2022/06/02 2022, doi: 10.1155/2022/6473507.

\bibitem{Nimbalkar21}
P. Nimbalkar and D. Kshirsagar, “Feature selection for intrusion detection system in Internet-of-Things (IoT),” ICT Express, vol. 7, no. 2, 2021, pp. 177-181.

\bibitem{Hindy}
H. Hindy, E. Bayne, M. Bures, R. Atkinson, C. Tachtatzis, and X. Bellekens, “Machine learning based IOT intrusion detection system: An MQTT case study (MQTT-IOT-IDS2020 dataset),” Selected Papers from the 12th International Networking Conference, pp. 73–84, 2021.

\bibitem{Koroniotis}
N. Koroniotis, N. Moustafa, E. Sitnikova, and B. Turnbull, “Towards the development of realistic botnet dataset in the Internet of Things for network forensic analytics: Bot-IoT dataset,” Future Generation Computer Systems, vol. 100, 2019, pp. 779-796.

\bibitem{WUSTL-EHMS}
Raj Jain, “WUSTL EHMS 2020 Dataset for Internet of Medical Things (IoMT) Cybersecurity Research,” 2020. [Online]. Available: https://www.cse.wustl.edu/~jain/ehms/index.html. [Accessed: 03-Oct-2023]. 

\bibitem{MQTT-IoT}
H. Hindy, C. Tachtatzis, R. Atkinson, E. Bayne, X. Bellekens, June 23, 2020, “MQTT-IoT-IDS2020: MQTT Internet of Things Intrusion Detection Dataset,” IEEE Dataport.

\bibitem{BOT-IoT19}
N. Moustafa, October 16, 2019, “The Bot-IoT dataset,” IEEE Dataport.

%\bibitem{DT21}
%Wikipedia, “Decision tree learning,” Wikipedia, 31-Dec-2021. [Online]. Available: https://en.wikipedia.org/wiki/Decision\_tree\_learning. [Accessed: 21-Jan-2022]. 

%\bibitem{RF08}
%J. Zhang, M. Zulkernine, and A. Haque, “Random-forests-basednetwork intrusion detection systems,” IEEE Transactions on Systems, Man, and Cybernetics, Part C (Applications and Reviews), vol. 38, no. 5, pp. 649– 659, 2008.

%\bibitem{ANN22}
%Wikipedia, “Artificial neural network,” Wikipedia, 16-Jan-2022. [Online]. Available: https://en.wikipedia.org/wiki/Artificial\_neural\_network. [Accessed: 21-Jan-2022].

\bibitem{sklearnDT}
scikit-learn "DecisionTreeClassifier,” 2023. [Online]. Available: https://scikit-learn.org/stable/modules/generated/sklearn.tree.\\DecisionTreeClassifier.html. [Accessed: 03-Oct-2023]. 

\bibitem{sklearnRF}
scikit-learn, “RandomForestClassifier,” 2023. [Online]. Available: https://scikit-learn.org/stable/modules/generated/sklearn.ensemble.\\RandomForestClassifier.html. [Accessed: 03-Oct-2023]. 

\bibitem{kerasANN}
Keras, “The Sequential model,” 2020. [Online]. Available: https://keras.io/guides/sequential\_model/. [Accessed: 03-Oct-2023]. 

%\bibitem{kfold}
%Wikipedia, “Cross-validation (statistics),” Wikipedia, 9-Jan-2022. [Online]. Available: https://en.wikipedia.org/wiki/Cross-validation\_(statistics). [Accessed: 21-Jan-2022].

%\bibitem{f1-score}
%Wikipedia, “F-score,” Wikipedia, 10-Jan-2022. [Online]. Available: https://en.wikipedia.org/wiki/F-score. [Accessed: 21-Jan-2022].

\bibitem{safety-score}
T. Salman, A. Ghubaish, D. Unal, and R. Jain, “Safety Score as an Evaluation Metric for Machine Learning Models of Security Applications,” in IEEE Networking Letters, vol. 2, no. 4, pp. 207-211, 2020.

\end{thebibliography}
\end{document}